\documentclass{ws-procs9x6}
\usepackage{psfig}

\def\etal{{\it et~al.\ }}
\def\la{\ifmmode\stackrel{<}{_{\sim}}\else$\stackrel{<}{_{\sim}}$\fi} 
\def\ga{\ifmmode\stackrel{>}{_{\sim}}\else$\stackrel{>}{_{\sim}}$\fi} 
\def\kms{km~s$^{-1}$}

\begin{document}

\title{Shocks, Outflows and Bubbles: \\
New Views on Pulsars and their Winds}

\author{BRYAN~M. GAENSLER}

\address{Harvard-Smithsonian Center for Astrophysics, \\
60 Garden Street MS-6, \\
Cambridge MA 02138, USA \\
E-mail: bgaensler@cfa.harvard.edu}


\maketitle

\abstracts{ A typical young pulsar slows down at an imperceptible rate,
its spin period increasing by less than 10 microseconds over the course
of a year.  However, the inertia of a pulsar is so extreme that to
effect this tiny change in rotation rate, the star must dissipate about
$10^{46}$~ergs of kinetic energy.  Observations of pulsars and their
surroundings demonstrate that this ``spin-down energy'' is expelled
into the pulsar's surroundings in spectacular fashion, in the form of a
relativistic wind of charged particles and magnetic fields.  In this
review I highlight some recent observational results on pulsar winds at
radio, X-ray and optical wavelengths, and explain what we can learn
from these data about shock structure, particle acceleration and the
interstellar medium. }

\section{Introduction}

Pulsars are thought to be born with very rapid initial periods,
typically 10--100~ms (see Migliazzo \etal\ 2002). 
The rotational energy of a new-born
neutron star can thus be very large, potentially approaching the
$\sim10^{51}$~ergs of kinetic energy released in the surrounding
supernova explosion. Observationally, it can be easily established that
the spacing between the regular radio pulses seen from these
sources is steadily increasing, demonstrating that all young pulsars
are slowing down. This ``spin-down'' implies
that a pulsar's enormous reservoir of kinetic energy is
being dissipated at a furious rate ---
for the youngest sources we find
$\dot{E} \equiv I \omega \dot{\omega} \approx 10^{35} - 10^{39}$~erg~s$^{-1}$,
where $\omega$ is a pulsar's angular frequency of rotation,
$I \equiv 10^{45}$~g~cm$^2$ is the moment of inertia usually
adopted for such sources, and $\dot{E}$ is the pulsar's
``spin-down luminosity''.

Ultimately this energy dissipation is due to the torques produced by
extreme electromagnetic fields on and near the surface of the neutron
star; indeed the observed values of $\omega$ and $\dot{\omega}$ are
routinely used to infer a neutron star's surface magnetic field
(Manchester \& Taylor 1977). Further
from the star, the bulk of the energy associated with
these torques goes into a relativistic ($\gamma \approx 10^6$)
magnetized wind (Michel 1969), through which the star ultimately
deposits its rotational energy into its surroundings.  (Only a tiny
fraction of this spin-down luminosity goes into generating the coherent
emission from the magnetic poles which results in the star's
characteristic pulsations.)

When a relativistic pulsar wind is confined by external pressure, this
source become directly observable as a {\em pulsar wind
nebulae}\ (PWN)\footnote{sometimes also referred to as a ``plerion''},
of which the Crab Nebula is the most spectacular example.  These
sources are our best guide to the processes through which a pulsar
loses its energy and couples to its surroundings.  Because PWNe are
close enough to be spatially resolved, they provide a wonderful
laboratory for directly observing the physics of particle acceleration
and strong shocks, processes which manifest themselves in a variety of
other situations in high-energy astrophysics.

In this review, I summarize what we know about
the overall properties and detailed structure of PWNe,
with particular emphasis on the flood of recent results
which have revitalized this field. 

\section{Pulsar Wind Nebulae}

Broadly speaking, there are three types of PWN, reflecting the
evolutionary state of the system and the ambient conditions (see 
Chevalier 1998 for a more detailed discussion).

A very young ($\sim1000$~yr) pulsar is typically embedded inside an
equally young supernova remnant (SNR). The pulsar's wind is
significantly over-pressured with respect to the cold ejecta inside the
SNR shell. The PWN thus expands rapidly in all directions: one
typically observes a quasi-spherical radio and X-ray nebula, with the
pulsar at its center. The Crab Nebula, as shown in
Figure~\ref{fig_bubble}, is the prototype of such sources (although in
this case the surrounding SNR has not yet been directly detected).  I and others
have in the past referred to these sources as ``static PWNe'', to
differentiate them from bow-shock nebulae as described in
\S\ref{sec_bow} below. However,
given that these sources are rapidly expanding, this terminology can be
confusing --- here I prefer to call these sources ``bubble''
PWNe, to reflect the fact that (deviations from spherical symmetry
aside) these sources are simply expanding bubbles.

At later stages in evolution ($\ga10\,000$~yr), the SNR
evolves towards the ``Sedov-Taylor'' phase of evolution,
in which a reverse shock fills the SNR cavity with
shocked ejecta. The collision between this reverse
shock and the expanding PWN has a dramatic effect,
crushing and compressing the pulsar nebula (e.g.\ Blondin \etal\ 2001).
The resulting PWN can have drastically different
appearances in the radio and X-ray bands, with the pulsar
substantially offset from the center of the PWN,
and the PWN offset from the center of the SNR.

At much later times, the SNR has dissipated and the pulsar is now
moving at a space velocity of typically $\sim500$~\kms\ through ambient
gas.  The ram pressure of the pulsar's wind generates a bow shock,
which can be seen as a cometary structure in radio and X-rays,
and also possibly
in H$\alpha$ and other optical lines.  PSR~B1957+20, shown in
Figure~\ref{fig_bow}, is a beautiful example of such a system.

\subsection{Bubble Pulsar Wind Nebulae: Overall Properties}

``Bubble'' PWNe, as typified by the Crab Nebula, characteristically
have a centrally filled morphology at all wavebands and show
significant levels of linear polarization. Furthermore, many of these
sources are increasingly smaller in spatial extent at progressively
higher energies.  These observations provide convincing evidence that
it is synchrotron emission which we detect from these sources,
generated by relativistic electrons and positrons supplied by the
central pulsar.

In many of these sources, the synchrotron lifetimes of emitting
particles at radio and X-ray energies are significantly more than and
less than the age of the source, respectively. This implies that the
radio luminosity of such a PWN traces the integrated history of the
pulsar's spin-down since birth, while the X-ray luminosity traces the
pulsar's current behavior. When combined, these data can be used to
build up a broadly consistent picture for the pulsar's evolution and current
properties.

The broadband spectrum of such PWNe is inevitably more complicated than a
single power law, however, generally showing multiple spectral
breaks.
Such spectral breaks are often attributed to synchrotron
losses, from which the nebular magnetic field can then be inferred
(e.g.\ Frail \etal\ 1996).  
However, these spectral features can also be produced by sudden
changes in the pulsar's rate of spin-down or by intrinsic breaks in the
underlying particle spectrum (Pacini \& Salvati 1973;
Woltjer \etal\ 1997).  We currently do not have a good
understanding of how to interpret the broadband spectral features of
PWNe, and caution should therefore be exercised in interpreting them.

\subsection{Bubble Pulsar Wind Nebulae: Structure}

\begin{figure}
\centerline{\psfig{file=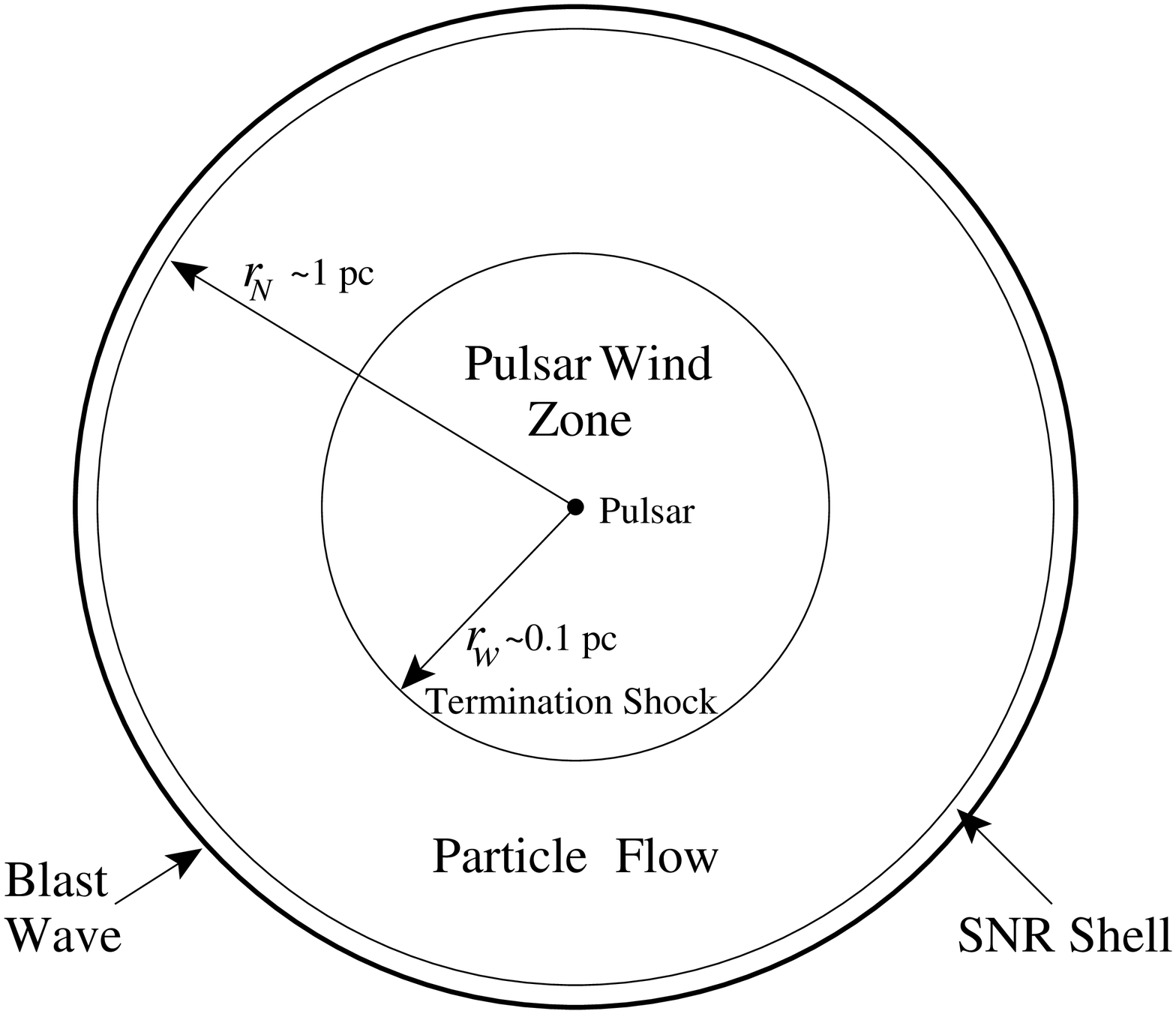,height=5cm}
\hspace{0.025\textwidth}
\psfig{file=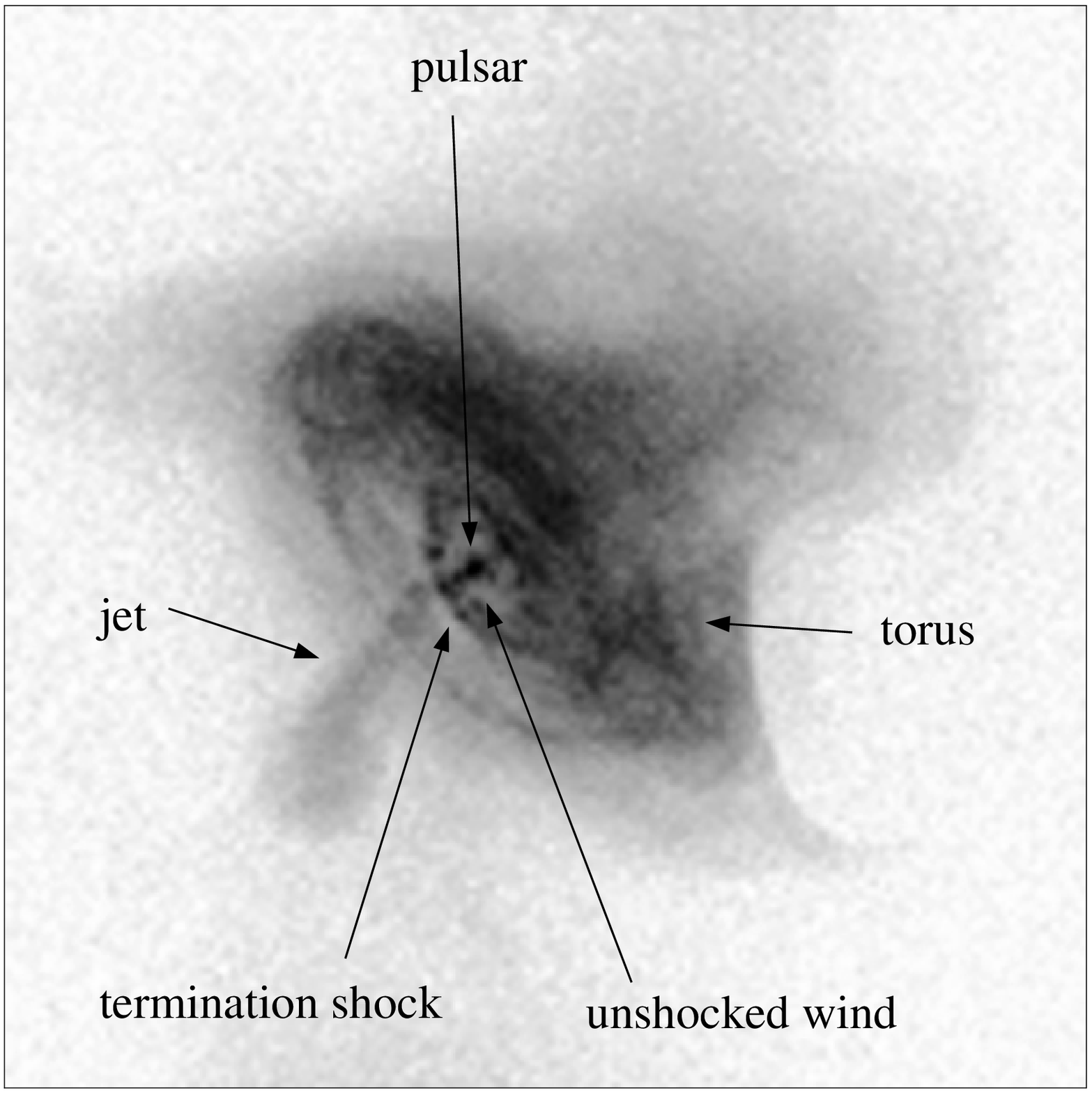,height=5cm}}
\caption{Cartoon of a bubble PWN (left; Slane 2002), compared to
a {\em Chandra}\ X-ray image of the Crab Nebula (right; Weisskopf \etal\ 2002).}
\label{fig_bubble}
\end{figure}

At the most basic level, a bubble PWNe can be simply approximated as
consisting of three zones, as shown in Figure~\ref{fig_bubble}. 
In the inner regions
near the pulsar, a relativistic but cold particle wind is launched from
the pulsar light cylinder. This wind is not directly observable.

At some distance $r_w$ from the pulsar, the pressure of the outflowing
pulsar wind is balanced by gas and magnetic pressure in the surrounding
emitting region. At this {\em termination shock}, particles are
thermalized and accelerated up to X-ray emitting energies.  
Downstream of the termination shock, the flow continues
to decelerate, and generates observable synchrotron emission.

While there is considerable support for the validity of
this simple picture, until recently only the
outer, emitting, region of most PWNe had been identified. There have
been remarkable advances in this regard with the advent of the {\em
Chandra X-ray Observatory}, whose unparalleled spatial resolution has
allowed us to delineate the termination shock and the region within it
in numerous PWNe, as shown in Figure~\ref{fig_bubble}. Because the termination
shock marks the point at which particles are accelerated and the
wind shocks, data on this region are our best handle yet
on the process by which the pulsar wind couples with and
reacts to its environment. 

{\em Chandra}\ has demonstrated that in many PWNe the termination shock
clearly has a toroidal, rather than spherical, geometry, suggesting
that the pulsar wind is focused into an equatorial flow, as is indeed
expected for an oblique rotator (e.g.\ Coroniti 1990). In this
interpretation, the orientation and eccentricity of this ring allow us
to simply determine the orientation of the pulsar's spin axis in three
dimensions. This information, when combined with measurements of pulsar
proper motions and pulse polarization, can provide unique insight into
pulsar emission mechanisms and neutron star formation (Helfand
\etal\ 2001; Lai  \etal\
2001).  Significant brightness variations can be seen around the
perimeters of these termination shocks. If interpreted as resulting
from Doppler boosting, these brightness variations yield the velocity
of the post-shock flow, which in turn can be used to infer the ratio of
magnetic to particle energy in the shocked wind (Lu \etal\ 2002).  In
some sources, the termination shock region shows a complex morphology,
comprised of multiple concentric wisps or arcs; this may
represent internal shock structure, as can be produced by gyrating ions
in the wind (Gaensler \etal\ 2002a).  {\em Chandra}\ observations are also
now beginning to reveal remarkable time-variability
in these regions, demonstrating the highly
dynamic nature of the relativistic flow (e.g.\ Hester \etal\ 2002).
Clearly all these various compact features provide
our best diagnostic yet of the nebular energetics and structure.

While results from {\em Chandra}\ have strikingly verified that bubble
PWNe indeed have the basic structure shown in Figure~\ref{fig_bubble}, these
observations have also  demonstrated that the assumptions of
isotropy and sphericity also shown in this Figure are highly
questionable. Most spectacular are the bright and collimated ``jets''
seen for several sources as shown in Figure~\ref{fig_bubble}, all oriented along the
symmetry axis defined by the inner toroidal termination shock. These
features have a wide range in brightnesses and extents, so one should
be cautious of applying a blanket interpretation to them.  However, the
fact that these structures have a symmetry which aligns with that seen
in the termination shock region argues that they lie (at least in
projection) along the pulsar spin axis. While some authors have argued
that these structures are not true jets (e.g.\ Radhakrishnan
\& Deshpande 2001), direct observations show
material flowing along these structures at $v \ga 0.3c$ 
(Hester \etal\ 2002), while the lack of synchrotron cooling inferred from their
spectra requires a similarly high flow speed (Gaensler \etal\ 2002a). 
In some cases these
jets are distinctly one-sided; it is not yet clear whether this
is due to Doppler boosting, or whether there is an intrinsic asymmetry
in these outflows. It is also not clear how these jets are formed or
collimated, but presumably the magnetic field of the neutron star plays
an important role (e.g.\ Lyubarsky \& Eichler 2001).

\subsection{Bow-Shock Pulsar Wind Nebulae}
\label{sec_bow}

When a pulsar is near the edge of or outside its associated SNR,
it is the ram pressure of the pulsar's motion which now serves to confine
the pulsar wind. The PWN consequently has a characteristic bow-shock
morphology, whose scale is set by the stand-off
distance, $R_0$, between the pulsar and the apex of the bow shock.
Pressure balance implies:
\begin{equation}
\frac{\dot{E}}{\Omega r_w^2 c} \approx \rho V^2 ,
\end{equation}
where $\Omega$ is the solid angle through which the wind flows,
$\rho$ is the ambient density, and $V$ is the pulsar's space velocity;
$R_0$ typically differs from the termination 
shock radius, $r_w$, by a numerical constant of order 1 (Bucciantini 2002).
For warm neutral regions of the
interstellar medium ($\rho\sim 1\times10^{-24}$~g~cm$^{-3}$), the corresponding
angular scale is $\sim1''$. 

\begin{figure}
\centerline{\psfig{file=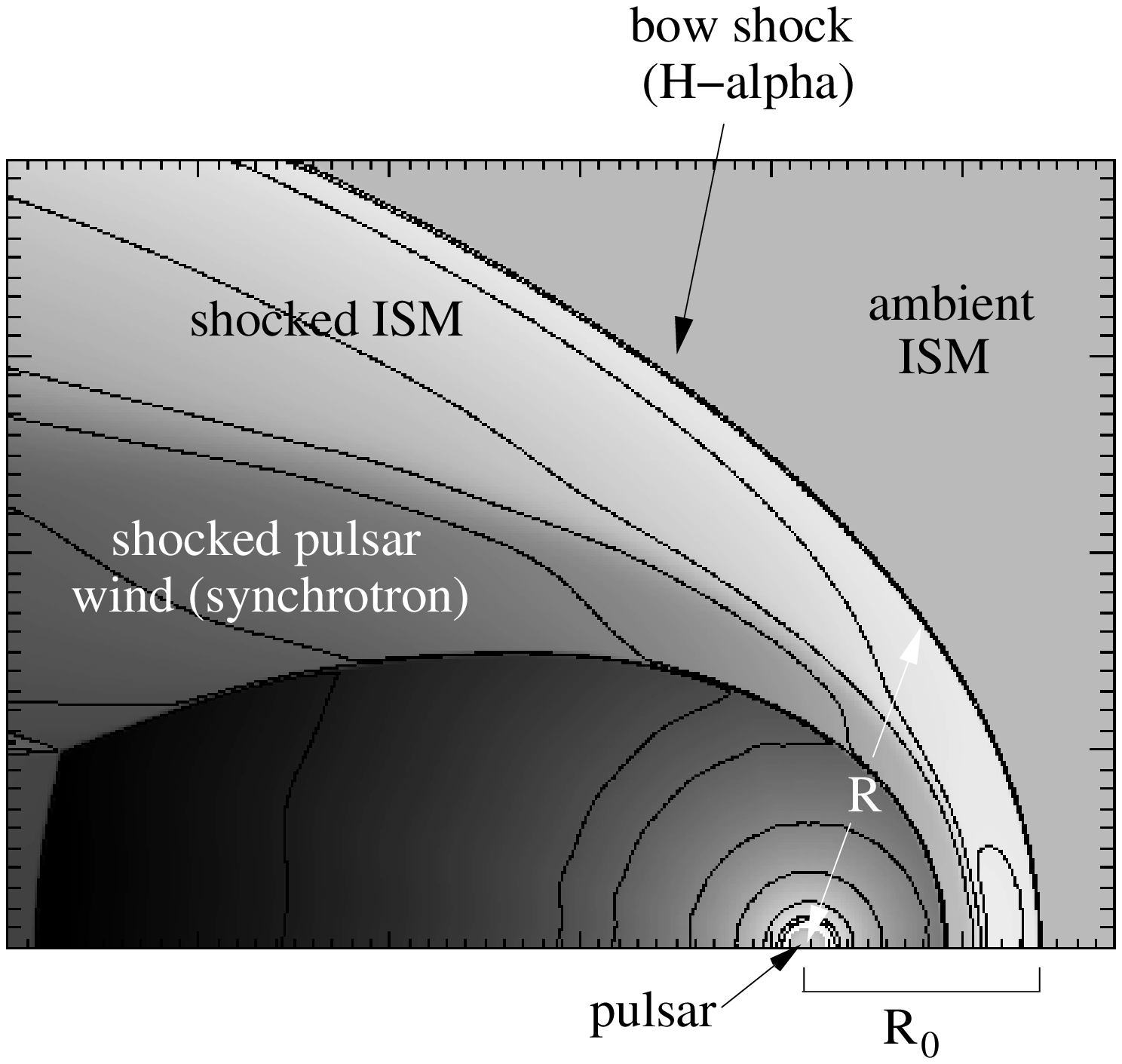,height=5cm,clip=}
\hspace{0.025\textwidth}
\psfig{file=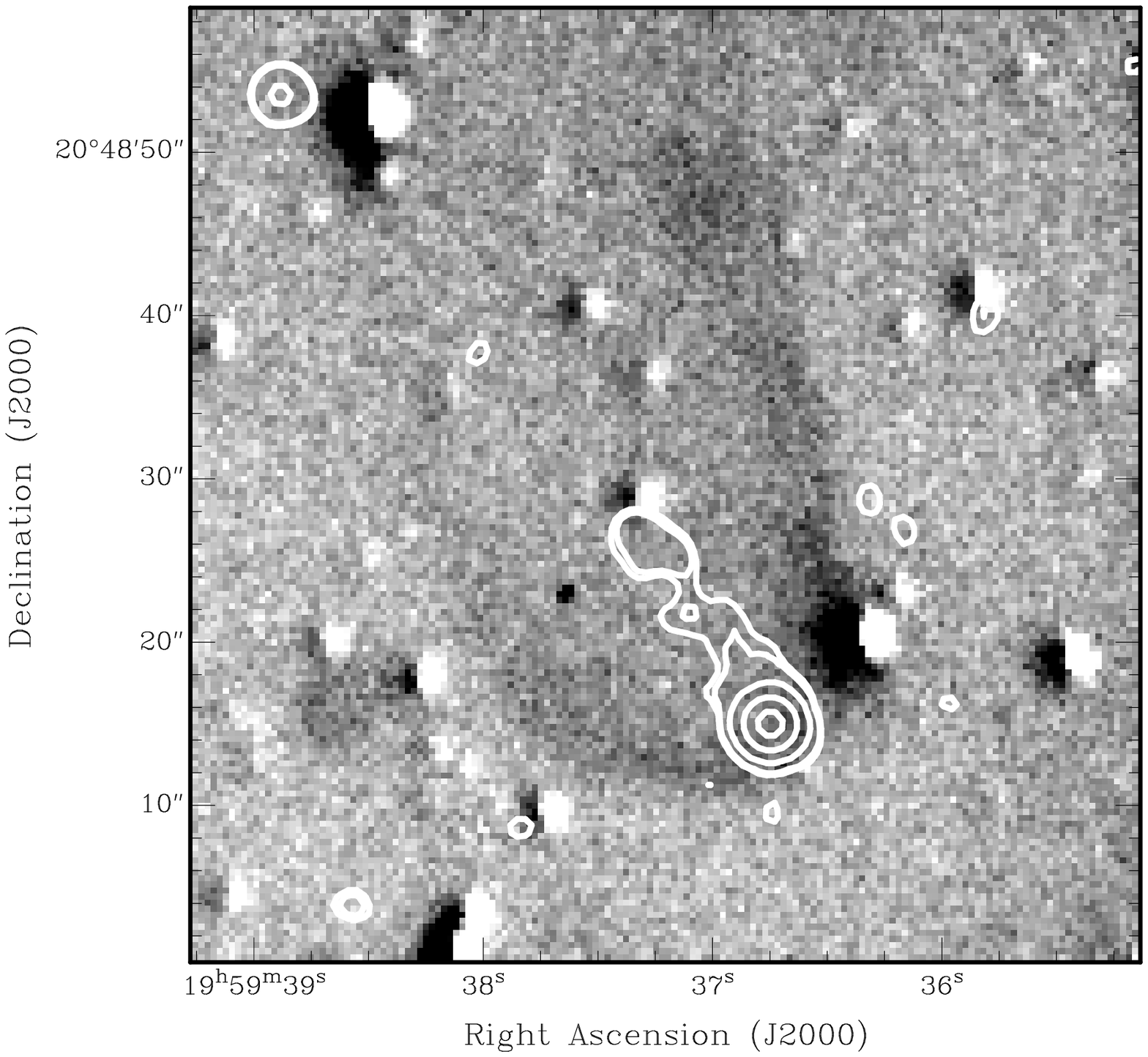,height=5cm}}
\caption{Cartoon of a bow-shock PWN (left; Bucciantini 2002), compared to
an optical (greyscale; AAT) plus X-ray (contours; {\em Chandra}) observation
of PSR~B1957+20 (right; Stappers \etal\ 2003).}
\label{fig_bow}
\end{figure}

The overall structure of a bow shock is shown in Figure~\ref{fig_bow}.
Just as for a bubble PWN, the nebula can be divided up into zones
corresponding to the freely flowing wind, the shocked wind, the shocked
ISM and ambient undisturbed material. In this case two separate
emission mechanisms are potentially observable:  collisional excitation
of ambient neutrals at the outer shock can generate H$\alpha$ emission,
while the internal shocked pulsar wind should produce X-ray/radio
synchrotron emission. Until recently, there
was yet to be a pulsar in which both these regions had been
identified simultaneously,
but new optical/X-ray data on PSR~B1957+20 (Figure~\ref{fig_bow};
Stappers \etal\ 2003) now
wonderfully demonstrate both the outer optical bow shock and the inner
X-ray termination shock in the same source.

The morphology, velocity structure and brightness of a bow shock
in the H$\alpha$ line provide powerful diagnostics
of the properties of the pulsar wind and of its interaction
with the ISM. Most simply, in the idealized situation of
an isotropic wind interacting with a homogeneous ambient medium,
the shape of a bow shock should have a
simple analytic form (Wilkin 1996):
\begin{equation}
R(\theta) = R_0 / \sin \theta \sqrt{3(1-\theta/\tan \theta)},
\label{eqn_bow}
\end{equation}
where $R$ is the distance
of the bow shock from the pulsar at an angle $\theta$ from the apex.
In some cases (e.g.\ PSR~J0437--4715; Fruchter 1995), Equation~(\ref{eqn_bow})
provides a good match to the observed morphology; however
in other sources (e.g.\ PSR~J2124--3358; Gaensler \etal\ 2002b), the nebula
deviates strongly from this expectation, implying anisotropies in the
pulsar wind and/or structure in the ISM.
Additional information is starting to become 
available by studying the time evolution
of H$\alpha$ bow shocks over time-scales of $>$5--10~years. The
observed changes in shape and brightness as a function of time
provide a unique probe
of density fluctuations in the ISM on scales of $\sim1000$~AU.

\section{Conclusions}

Interest in pulsar nebulae has resurged in the last few years,
because of a dramatic increase in the quality and quantity of
data across the electromagnetic spectrum. We can now be confident
that we are seeing in our data the predicted shock structures for both
bubble and bow-shock nebulae, and can consequently use the geometry
of a PWN to infer the orientation of the pulsar's spin axis,
and the magnitude and direction of the pulsar's motion. Spectral,
spatial and temporal variations in the emission from PWNe now provide
numerous probes of the way in which a pulsar interacts with its surroundings.

Many questions still remain unanswered. How are the 
prominent jets seen in several sources formed and collimated?
What are the physical processes which account for the complicated
compact structures and remarkable time variability now being
identified in many sources?
And ultimately, what are the drivers behind the diversity
seen in PWN morphologies? While these are difficult questions to
answer, they are now questions which we have the 
data to actually address.  Clearly we have entered a new
era in the study of pulsars and their winds.

\section*{Acknowledgments}
I acknowledge the support
of NASA through SAO grants
GO2-3041X, GO2-3075X and GO2-3079X, and through
XMM-Newton Guest Observer grant
NAG5-11376.

\end{document}